\documentclass[12pt]{article}
\usepackage[margin=2cm]{geometry}
\usepackage{comment}
\usepackage{amsmath,amssymb,extarrows,mathtools,graphicx,subfigure,setspace}
\usepackage{cite}
\usepackage{slashed}
\usepackage{color}
\makeatother

\newcommand{\be}{\begin{equation}}
\newcommand{\bea}{\begin{eqnarray}}
\newcommand{\eea}{\end{eqnarray}}
\newcommand{\ba}{\begin{array}}
\newcommand{\ea}{\end{array}}
\newcommand{\ee}{\end{equation}}
\newcommand{\bes}{\begin{equation*}}
\newcommand{\beas}{\begin{eqnarray*}}
\newcommand{\eeas}{\end{eqnarray*}}
\newcommand{\bas}{\begin{array*}}
\newcommand{\eas}{\end{array*}}
\newcommand{\ees}{\end{equation*}}

\numberwithin{equation}{section}
\begin{document}
\onehalfspacing
\vfill
\begin{titlepage}
\vspace{10mm}
\begin{flushright}
 IPM/P-2014/002 \\
\end{flushright}
  %
\vspace*{20mm}
\begin{center}
{\Large {\bf  Thermalization in backgrounds with hyperscaling violating factor     }\\
}

\vspace*{15mm}
\vspace*{1mm}
{Mohsen Alishahiha${}^a$, Amin Faraji Astaneh$^{b,c}$ 
and  M. Reza Mohammadi Mozaffar$^a$ }

 \vspace*{1cm}

{\it ${}^a$ School of Physics, Institute for Research in Fundamental Sciences (IPM),\\
P.O. Box 19395-5531, Tehran, Iran \\  
${}^b$ Department of Physics, Sharif University of Technology,\\
P.O. Box 11365-9161, Tehran, Iran\\
${}^c $ School of Particles and Accelerators,\\Institute for Research in Fundamental Sciences (IPM)\\
P.O. Box 19395-5531, Tehran, Iran \\ 
}
 \vspace*{0.5cm}
{E-mails: {\tt alishah@ipm.ir, faraji@ipm.ir, m$_{-}$mohammadi@ipm.ir}}%

\vspace*{1cm}
\end{center}

\begin{abstract}
We present an  analytic solution of a Vaidya-charged black hole with a hyperscaling
 violating factor in 
an Einstein-Maxwell-dilaton model, where the scalar potential plays a key role in the existence of the solution.  
By making use of this result, we  study the process of thermalization  after 
a global quench in a theory which its gravitational description is provided by the resultant 
solution in the case of zero charge. In particular, we  probe the system by entanglement entropy and show that 
it exhibits certain scaling behaviors during the process.

\end{abstract}

\end{titlepage}

\section{INTRODUCTION}

One of the interesting phenomena which appears  in different areas of physics is the
process of thermalization of a nonequilibrium state. In the process of thermalization, there are several interesting issues which are worth exploring. These  include, for example,  how fast the process of thermalization is and what kind of quantities can probe  the process.
Indeed, being out of equilibrium the thermodynamical quantities such as temperature, entropy, 
pressure, etc.,  may not be well defined during  the process of thermalization. Therefore, 
the standard thermodynamics which usually provides a useful tool
to study long-range physics is not applicable in this case.
Nevertheless, there might be other quantities that could probe the system when it thermalizes from a 
nonequilibrium state to a thermal state . A prototype example of such quantities is 
entanglement entropy, which can be defined even when the system is out of equilibrium.

Of course, for a generic quantum system it is difficult to compute entanglement entropy,
though for those systems (typically the strongly coupled ones) which have gravitational 
descriptions\cite{Maldacena:1997} one may utilize the holographic description of the entanglement entropy\cite{{RT:2006PRL},{RT:2006}}
to study its  behavior. We note, however, that since the
system is time dependent in order to compute the holographic entanglement entropy one should use its covariant proposal\cite{Hubeny:2007xt}.

In the context of gauge/gravity duality the process of thermalization  
 may be mapped to a black hole formation due to a gravitational collapse
which can be modelled  by  an AdS-Vaidya metric.
The corresponding metric describing  the collapse of neutral matter  in $D+2$ dimensions is
\be\label{m0}
 ds^2=\frac{1}{\rho^2}\bigg(f(\rho,v)dv^2-2 d\rho dv+d\vec{x}^2\bigg),\;\;\;\;\;\;\;\;\;
f(\rho,v)=1-m(v)\;\rho^{D+1},
\ee
 where $\rho$ is the radial coordinate, $x_i$'s $(i=1,... ,D)$ are spatial boundary coordinates, and $v$ is the null coordinate.  Here $m(v)$ is an arbitrary function of the null coordinate $v$ satisfying 
$\partial_v m(v)>0$. This constraint is required by the null energy condition. In what follows, 
we will consider the case of $m(v)=m \;\theta(v)$, where $m$ is a constant and $\theta(v)$
 is the step function. Therefore for $v<0$ the geometry is an AdS metric, while
for $v>0$ it is an AdS-Schwarzschild black hole whose  horizon is located at $\rho_H=m^{-1/(D+1)}$.
The AdS radius is also set to one.

The above background (with $m(v)=m \;\theta(v)$) could provide a gravitational description for 
a sudden change in a strongly coupled field theory  which might be caused by  turning on a  source for an operator in an  interval $\delta t\rightarrow 0$.
This change can excite the system to an excited state with nonzero energy density that could
eventually thermalize to an equilibrium thermal state. Since we are considering a sudden change in the
theory,  it is then natural to think of the  process  as a thermalization after {\it a global quantum quench}.

Therefore, in order to study  entanglement entropy during the process of thermalization, after 
a global quantum quench  one needs to compute  the holographic entanglement entropy in the above time-dependent background by using its covariant proposal\cite{Hubeny:2007xt}.
Holographically, the entanglement entropy can be computed by  extremizing the area of a codimension-two  hypersurface in the  AdS-Vaidya geometry \eqref{m0} whose boundary coincides with the 
boundary of the entangling region\cite{{RT:2006PRL},{RT:2006},{Hubeny:2007xt}}.
Indeed, by making use of this prescription, the holographic entanglement entropy 
in this background has been 
studied in several papers,  including Refs. \cite{{AbajoArrastia:2010yt},{Albash:2010mv},{Balasubramanian:2010ce},{Aparicio:2011zy},{Galante:2012pv},{Caceres:2012em},{Baron:2012fv},{Fischler:2012ca},{Fischler:2012uv},{Li:2013sia},{Shenker:2013pqa},{Caputa:2013eka},{Fischler:2013fba}}, where it was shown that 
the holographic entanglement entropy exhibits different scaling behaviors as the system evolves in time.

The above consideration may be compared with the results of  Ref. \cite{CC},  where the behavior 
of the entanglement entropy after a  global quantum quench for a two-dimension CFT was studied.
Although the quench considered in this case is  different from that studied holographically,
there is an agreement between the results of these two different setups. Of course, this agreement 
might be 
understood from the fact that in both cases one is considering  the evolution of an excited state
in a CFT. Keeping this  distinction in our mind, in what follows we will refer to  our setup as a gravitational description for a thermalization process after a global quantum  quench.

Generally, the behavior of the entanglement entropy during the evolution of the system 
consists of two phases: a time growing phase and a saturation phase where the entanglement 
entropy saturates to its equilibrium value. Actually, one may associate two time scales to the system which
are  the radius of the horizon, $\rho_H$, and 
the size of the entangling region. For example, if the  entangling region is a strip, the size is given by 
its  width $\ell$. Indeed, the theory would reach local equilibrium at $t\sim \rho_H$ when
the system stops producing thermodynamic entropy, while at $t\sim \frac{\ell}{2}$ the entanglement 
entropy saturates to its equilibrium value \cite{{AbajoArrastia:2010yt},{Albash:2010mv},{Balasubramanian:2010ce}}. When $\frac{\ell}{2}\lesssim \rho_H$
the entanglement entropy saturates at $t\sim \frac{\ell}{2}$ before the system reaches local equilibrium, whereas for $\frac{\ell}{2}\gg \rho_H$ the entanglement entropy is far from 
its equilibrium value even though the system is locally equilibrated.

From a gravity point of view, when  $\frac{\ell}{2}\lesssim\rho_H$ the corresponding codimension-two
hypersurface is always outside the horizon,\footnote{Actually, it was observed in Ref. 
\cite{Hubeny:2012ry} that 
the extremal surfaces in the bulk cannot penetrate through the horizon of a static black hole.} whereas for $\frac{\ell}{2}\gg \rho_H$  the codimension-two hypersurface may penetrate into the horizon. Actually, in this case 
the evolution of the entanglement entropy is  controlled by the geometry around and inside the horizon
 for  $t\gtrsim \rho_H$ \cite{{Liu:2013iza},{Liu:2013qca}}.

The aim of this paper is to extend the consideration of Refs. \cite{{Liu:2013iza},{Liu:2013qca}}  for those
strongly coupled theories whose gravitational descriptions are provided by a Vaidya metric with a hyperscaling 
violating factor\cite{{Charmousis:2010zz},{Gouteraux:2011ce},{Dong:2012se}}.\footnote{A relativistic
 system
with a hyperscaling violation was also discussed in Refs. \cite{{Liu:2013iza},{Liu:2013qca}}. } To do so, we first need to find a Vaidya metric with a
hyperscaling violating factor. This is, indeed, what we will do in the next section. Then we will study the time dependence of holographic entanglement entropy in the resultant geometry. The entanglement entropy for Vaidya-Lifshitz geometry has been also studied in \cite{Keranen:2011xs}.

Although our original intention was to study the behavior of  entanglement entropy during 
a global quantum quench in a strongly coupled field theory whose gravitational
description is provided by the Vaidya metric with a
hyperscaling violating factor, one could also study other quantities such as the Wilson loop and 
equal-time two-point function by making use of gauge/gravity duality. Indeed, in these cases
one still needs to extremize the area of certain hypersurfaces in our resultant background.
In fact, for the Wilson loop it is a two-dimensional surface\cite{{Rey:1998ik},{Maldacena:1998im}}, 
while for  the equal-time two-point function it is just a geodesic\cite{Balasubramanian:1999zv}.
The behaviors of the Wilson loop and equal-time two-point function of an operator in a global
quantum quench in a field theory whose gravity dual is the AdS-Vaidya metric \eqref{m0} have 
been studied in several papers including Refs.
\cite{{CC},{AbajoArrastia:2010yt},{Albash:2010mv},{Balasubramanian:2010ce},{Ebrahim:2010ra}}
(see also \cite{CC2}).

The paper is organized as follows. In the next section, we will find an analytic solution of an Einstein-Maxwell-dilaton model representing a Vaidya metric with a hyperscaling violating factor. In section three
we will consider holographic  entanglement entropy where we will also set up a formalism to study the 
Wilson loop and equal-time two-point function of an operator with  a large conformal dimension. 
In section four we will explore general behaviors of entanglement entropy when
the system evolves in time. In particular, we will consider entanglement entropy in more
detail for the case where the entangling region is much larger  than the radius of the horizon.
The last section is devoted to conclusions. Holographic entanglement 
entropy in a static black hole with a  hyperscaling violating factor is reviewed in Appendix A. In Appendix
B, some details of the computations are presented .

\vspace*{0.5cm}
{\bf Note added:} Few days after submitting our paper to  arXiv, \cite{Fonda:2014ula} appeared
where the same question has been addressed.  Indeed this paper has  a
substantial overlap with of ours. Of course,  the authors of  \cite{Fonda:2014ula} have considered 
both strip and sphere as entangling regions and also presented a numerical analysis. Where overlap exists, the results are in agreement.



\section{ INFALLING SHELL SOLUTIONS}

Hyperscaling violating geometries with nonzero charge have been studied in Ref.
\cite{alishahiha:2012}, where it was shown that an Einstein-Maxwell-dilaton model with
a particular potential admits such  solutions. The corresponding action is
\be
\label{action}
S=-\frac{1}{16\pi G_N}\int d^{D+2}x\sqrt{-g}\left[R-\frac{1}{2}(\partial\phi)^2+V_0 e^{\gamma \phi}-\frac{1}{4}
\sum_{i=1}^{N_g} e^{\lambda_i\phi}{F^{(i)}}^2\right], 
\ee
where  $G_N$ is the Newton constant,  $\gamma, V_0$, and 
$\lambda_i$ are free parameters of the model, and $N_g$ is the number of gauge fields.
One of the 
gauge fields is required  to produce an anisotropic
scaling symmetry, and  the above particular form of the  potential  is needed to be responsible to get a hyperscaling violating factor. The other gauge fields make the background charged.
In what follows, we will consider only $N_g=2$, though its generalization to other $N_g$ is straightforward.
 In this case, the corresponding solution is\cite{alishahiha:2012}
\bea\label{solution}
&&ds^2=r^{-2\frac{\theta}{D}}\bigg(-r^{2z}f(r)dt^2+\frac{dr^2}{r^2f(r)}+r^2d\vec{x}^2\bigg),
\;\;\;\;\phi=\phi_0+\beta\ln r,\cr &&\cr
 &&A^{(1)}_{t}=\sqrt{\frac{2(z-1)}{D-\theta+z}}\; e^{\frac{D-1+\theta/D}{\beta}\phi_0}
\;r^{D-\theta+z},\cr &&\cr
&&A^{(2)}_t=\sqrt{\frac{2(D-\theta)}{D-\theta+z-2}}\;e^{-\frac{z-1-\theta/D}{\beta}\phi_0}\;\frac{Q}{r^{D-\theta+z-2}},
\eea
with $\beta=\sqrt{2(D-\theta)(z-1-\theta/D)}$ and
\be
f(r)=1-\frac{m}{r^{D-\theta+z}}+\frac{Q^2}{r^{2(D-\theta+z-1)}},
\ee
where $z$ is the dynamical exponent and $\theta$ is the hyperscaling violation exponent. The parameters of the action are also found: 
\bea\label{para}
&&\lambda_1= \frac{2\theta (D-1)-2D^2}{D\beta},\;
\lambda_2=
\frac{\beta}{D-\theta},\;\gamma= \frac{2\theta}{D\beta},\;
V_0=\frac{
(D-\theta+z-1 ) (D-\theta+z  )}{e^{\gamma\phi_0}}.
\eea 
This is indeed a charged black brane solution whose horizon radius $r_H$ is obtained by
setting $f(r_H) = 0$. It is useful to define an {\it effective dimension} $d=D-\theta+1$ by which
the function $f$, appearing in the metric component $g_{vv}$, reads
\be
f(r)=1-\frac{m}{r^{d-1+z}}+\frac{Q^2}{r^{2(d-2+z)}}.
\ee
In this notation, the Hawking temperature and the thermal entropy of the solution are, respectively, 
\be\label{thermal}
T=\frac{(d-1+z)r_H^z}{4\pi}\left(1-\frac{(d-3+z)}{d-1+z}\;
\frac{Q^2}{r_H^{2(d-2+z)}}\right),\;\;\;\;\;S_{BH}=\frac{V_{D} r_H^{d-1}}{4G_N}\equiv V_D {\cal S}_{BH},
\ee
where $V_{D}$ is the volume of the spatial directions, $x_i, i=1,\cdots,D$, and ${\cal S}_{BH}$ is the entropy
density.

In what follows, we would like to find a background representing an infalling shell of massless
and pressureless charged matter in a hyperscaling violating geometry. The corresponding 
geometry may be thought of as a Vaidya metric with a hyperscaling violating factor.
The Vaidya-Lifshitz geometry has been studied in Ref. \cite{Keranen:2011xs}.

We note that  a charged  Vaidya space-time is sourced 
by an energy-momentum tensor and a current density of a massless null charged matter.
Therefore, in order to get such a solution, one needs to add a proper extra matter field to the action
\eqref{action}. By doing so,  the equations of motion of the action \eqref{action}
should be modified as follows:
\bea\label{EOM1}
&&R_{\mu\nu}+\frac{V(\phi)}{D}g_{\mu\nu}=\frac{1}{2}\partial_\mu\phi\partial_\nu\phi
+\frac{1}{2}\sum_{i=1}^{2}e^{\lambda_i\phi}\left(F_{\mu}^{(i)\; \rho} F^{(i)}_{\rho\nu}-\frac{g_{\mu\nu}}{2D}{F^{(i)}}^2\right)+T_{\mu\nu},\cr &&\cr
&& \nabla^2\phi=-\frac{dV(\phi)}{d\phi}+\frac{1}{4}\sum_{i=1}^{2}\lambda_i e^{\lambda_i\phi} 
{F^{(i)}}^2,\;\;\;\;\;\;
\nabla_\mu\left(\sqrt{-g}e^{\lambda_i\phi}F^{(i)\;\mu\nu}\right)=J^{(i)\;\nu},
\eea
where $T_{\mu\nu}$ and $J^{(i)\;\nu}$ are the energy-momentum tensor and current 
density of the charged matter field, respectively.
As we will see for the model we are considering, the 
corresponding nonzero components of the energy-momentum tensor and current 
density are $T_{vv}$ and $J^{(2)}_v$.

To find a charged Vaidya solution of the above equations, one may start with a proper ansatz for the 
metric and other fields. To do so, it is useful to introduce  an Eddington-Finkelstein-like  
coordinate system as follows:
\be
dv=dt+\frac{dr}{f(r)r^{z+1}},
\ee
by which the metric in Eq. \eqref{solution} may be recast to the following form:
\be
ds^2=r^{-2\frac{\theta}{D}}\bigg(-r^{2z}f(r)dv^2+2 r^{z-1}dr dv+r^2d\vec{x}^2\bigg).
\ee
Moreover, in this notation in the gauge of $A^{(i)}_{r}=0$ the nonzero  component of the
gauge field is $A^{(i)}_{v}$ and the dilaton remains unchanged. Motivated by this form of 
the metric,  let  us consider the following ansatz for the metric, scalar, and gauge fields:
\be\label{ansatz}
ds^2=r^{-2\frac{\theta}{D}}\bigg(-r^{2z}f(r,v)dv^2+2 r^{z-1}dr dv+r^2d\vec{x}^2\bigg),\;\;\;\;\phi=\phi(r),
\;\;\;\;\; F^{(1)}_{rv}(r)\neq 0,\;\;\;\;\;F^{(2)}_{rv}(r,v)\neq 0,
\ee
and all  other components of gauge fields are set to be zero. Note that 
in this ansatz the scalar field and the first gauge field, which are essential to 
support a solution with  an anisotropic scaling symmetry and hyperscaling violating factor, are independent of the null
coordinate $v$, whereas the function $f$, appearing in the metric component $g_{vv}$,  and the second gauge field
depend on both $r$ and $v$ coordinates.  

It is straightforward to plug this ansatz into the equations of motion \eqref{EOM1} to find the unknown functions.
Actually, from the $rr$ component of the Einstein equations, one finds 
\be
\phi=\phi_0+\sqrt{2(D-\theta)(z-1-\theta/D)}\ln r=\phi_0+\beta\ln r,
\ee
which is the same as that in the static case. In what follows, with no loss of generality one may set $\phi_0=0$.
Note that the null energy condition requires\cite{alishahiha:2012}
\bea\label{NEC}
(D-\theta)(z-1-\theta/D)\geq 0.
\eea
In this paper, we will consider the case of $z>1$ and $D>\theta$.
 
From the $ii$ components of the Einstein equations, taking into account the identifications of
 \eqref{para},
one can fix the function $f$ as follows:
\be
f(r,v)=1-\frac{m(v)}{r^{D-\theta+z}}+\frac{Q(v)^2}{r^{2(D-\theta+z-1)}},
\ee
where $m(v)$ and $Q(v)$ are arbitrary functions of $v$. The gauge fields can be also obtained from the
$v$ component of the Maxwell equations. Taking this equation into account, one can fix the solution completely:
\bea\label{solution2}
&&ds^2=r^{-2\frac{\theta}{D}}\bigg(-r^{2z}f(r,v)dv^2+2 r^{z-1}dr dv+r^2d\vec{x}^2\bigg),
\;\;\;\;\phi=\beta\ln r,\cr &&\cr
 &&A^{(1)}_{v}=\sqrt{\frac{2(z-1)}{D-\theta+z}}
\;r^{D-\theta+z},\;\;\;\;\;\;\;\;\;\;
A^{(2)}_v=\sqrt{\frac{2(D-\theta)}{D-\theta+z-2}}\;\frac{Q(v)}{r^{D-\theta+z-2}}.
\eea
 Of course, so far we have not used all equations of motion.  Indeed, from the $vv$ component 
of the Einstein equations  and $r$ component of the Maxwell equations, one may read the energy-momentum tensor and the current density which are needed to support an
infalling  shell solution.  More precisely, from the corresponding components of the 
equations of motion, one finds that the energy-momentum tensor and current density of the charged infalling  pressureless
matter are given by $T_{\mu\nu}={\varrho} U_\mu U_\nu$ and $J^{(2)}_\mu=\varrho_e U_\mu$ 
with $U_\mu=\delta_{\mu v}$, and
\bea
\varrho=\frac{\theta-D}{2}\frac{\partial f(r,v)}{\partial v}\; r^z,\;\;\;\;\;\;
\varrho_e=\frac{\partial Q(v)}{\partial v}\sqrt{2(D-\theta)(D-\theta+z-2)} \;r^{\theta-D}.
\eea
Note that the null energy condition requires $\varrho>0$.
Finally, the last nontrivial equation that needs to be checked is the $rv$ component of the Einstein equations. Actually, it is easy to see that this equation is also satisfied without imposing any further constraints.

The solution \eqref{solution2} can be thought of as a gravity solution which describes a gravitational 
collapse of charged matter in a model that has an anisotropic
scaling symmetry with a hyperscaling violating factor.
By using the gauge/gravity duality, it is plausible  to conjecture that this solution provides a gravity
description for the process of thermalization after a global quantum quench in a strongly coupled 
field theory with an anisotropic
scaling symmetry and violation of scaling. In what follows, we would like to probe this process by entanglement entropy (and also by the Wilson loop and equal-time two-point function of an operator with a large conformal dimension).


\section{ENTANGLEMENT ENTROPY}

Entanglement entropy may be considered as a useful quantity which could probe a system when
it undergoes a rapid change that generally brings the system out of equilibrium.  A global quantum quench is a prototype example of a rapid change. In this section, by making use of the 
holographic description of entanglement entropy, we will study the behavior of the entanglement 
entropy for a global quantum quench in a strongly coupled field theory  whose gravity dual is given by the solution \eqref{solution2}. 

To proceed, for simplicity, we set the charge to zero, $Q(v)=0$, so that the background is
neutral. In this case, by setting $r=\rho^{-1}$ the metric in Eq. \eqref{solution2} reads
\be\label{sol}
 ds^2=\rho^{2\frac{1-d}{D}}\bigg(-\rho^{2-2z}f(\rho,v)dv^2-2 \rho^{1-z}d\rho dv+d\vec{x}^2\bigg),
\;\;\;\;\;\;\;\;\;{\rm with}\;\;f=1-m(v)\rho^{d-1+z}.
\ee

Let us consider an entangling region in the shape of a strip with the width $\ell$ as follows:
\be
-\frac{\ell}{2}\leq x_1\equiv x\leq \frac{\ell}{2},\;\;\;\;\;\;\;\;\;0\leq x_a\leq L,\;\;\;\;\;{\rm for}\;\;a=2,\cdots,D.
\ee
Now the aim is to compute the entanglement entropy for this strip by using 
gauge/gravity duality. To do so, one should consider a codimension-two hypersurface in 
the geometry \eqref{sol}  whose boundary coincides with the boundary of the 
above strip. Of course, since the metric is not static, one needs to use the covariant proposal
for the holographic entanglement entropy\cite{Hubeny:2007xt}. Therefore,
the corresponding codimension-two hypersurface in the bulk 
may be  parametrized by $v(x)$ and $\rho(x)$. In fact, for the bulk metric \eqref{sol}, the induced metric 
 on the hypersurface is
\be
ds_{\rm ind}^2=\rho^{2\frac{1-d}{D}}\bigg[\bigg(1-\rho^{2-2z}f(\rho,v){v'}^2-2 \rho^{1-z} v'\rho'\bigg) dx^2+dx_a^2\bigg],
\ee
where the ``prime'' denotes derivative with respect to $x$. The area of the hypersurface is
\be\label{area00}
{\cal A}=\frac{L^{D-1}}{2}\int_{-\ell/2}^{\ell/2} dx\; \frac{\sqrt{1-2\rho^{1-z}v'\rho'-\rho^{2-2z} v'^2 f}}{\rho^{d-1}}\ .
\ee
Then the procedure is to extremize this area to read the entanglement entropy which is proportional to the area of the extremal hypersurface\cite{{RT:2006PRL},{RT:2006},{Hubeny:2007xt}}:
\be
S=\frac{\mathcal{A}_{d-1}}{4G_N}.
\ee
 It is worth 
noting that, although our main aim is to study  entanglement entropy during a global
quench after  which the system undergoes the process of thermalization, with a small
modification one may do even  more. Indeed, on top of  the entanglement entropy, we could also study the Wilson loop and equal-time two-point function of an operator with a large conformal dimension.
More precisely,  let us define the quantity ${\cal A}_n$ as follows:
\be\label{area0}
{\cal A}_n=\frac{L^{\frac{D}{d-1}n-1}}{2}\int_{-\ell/2}^{\ell/2} dx\; \frac{\sqrt{1-2\rho^{1-z}v'\rho'-\rho^{2-2z} v'^2 f}}{\rho^{n}}=\frac{L^{\frac{D}{d-1}n-1}}{2}\int_{-\ell/2}^{\ell/2} dx\; \frac{{\cal L}}{\rho^{n}}.
\ee
Then  entanglement entropy, the Wilson loop and the 
equal time two-point function of an operator with a large conformal dimension may be computed by 
extremizing ${\cal A}_n$ for different values of $n$. In particular, for entanglement entropy  one has  $n=d-1$, while for the  Wilson loop and  equal-time two-point function one should set 
$n=\frac{2(d-1)}{D}$ and  $n=\frac{d-1}{D}$, respectively.  Having extremized  ${\cal A}_n$, 
the corresponding quantities can be found as follows\cite{{Rey:1998ik},{Maldacena:1998im},{Balasubramanian:1999zv}}\footnote{Note that, in the case of 
a rectangular Wilson loop, $\ell$ is the width of the loop, and for the equal-time two-point function it is the distance between two operators in the boundary.}:
\bea\label{Qun}
&{\rm Wilson\;  loop}:\;\;\;\;\;\;\;\;\;\;& \langle W\rangle \sim e^{-\frac{\mathcal{A}_{{2(d-1)}/{D}}}{2\pi\alpha'}},\cr
&{\rm equal-time\; two-point\; function}:\;\;\;\;\;\;\;\;\;\;& G\sim e^{-M {\mathcal{A}_{{(d-1)}/{D}}}},
\eea
 where $(2\pi \alpha')^{-1}$ is the string tension and $M$ is the mass of the bulk field 
which is dual to the boundary operator whose  equal-time two-point function is computed. Although in what follows we 
will consider the entanglement entropy where $n=d-1$, we will keep working with unspecified
$n$ so that the final results may be also  used to read the Wilson loop and two-point function.

In order to extremize ${\cal A}_n$ we note that the expression \eqref{area0} may be thought of as
an action for a  one-dimensional dynamical system whose dynamical   fields are $v(x)$ and $\rho(x)$.
We note, however, that since the action is independent of $x$ the corresponding Hamiltonian is 
a constant of motion:
\bea\label{J}
H^{-1}=-\rho^{n}{\cal L}={\rm const.}
\eea 
Moreover, we have two equations of motion for $v$ and $\rho$. Indeed, by making use of the above conservation law, the  corresponding equations of motion are
\be\label{eom}
\partial_x P_v
=\frac{P_\rho^2}{2}\;\frac{\partial f}{\partial v},\;\;\;\;\;\;\;\;
\partial_x P_\rho=
\frac{P_\rho^2}{2}\;\frac{\partial f}{\partial \rho}+\frac{n}{\rho^{2n+1}}H^{-2}+\frac{1-z}{\rho^{2-z}}
P_\rho P_v,
\ee
where 
\be
P_v=\rho^{1-z}(\rho'+\rho^{1-z} v'f),\;\;\;\;\;\;\;\;\;\;\;\;P_\rho=\rho^{1-z}v',
\ee
are the momenta conjugate to $v$ and $\rho$ up to a factor of $H$, respectively. These equations have to be supplemented by  the following boundary conditions:
\bea\label{bdycondition}
\rho(\frac{\ell}{2})=0,\;\;\;\;\;\;\;\;v(\frac{\ell}{2})=t,\;\;\;\;\;\;\;\;\rho'(0)=0,\;\;\;\;\;\;\;\;v'(0)=0,
\eea
and 
\be
\rho(0)=\rho_t,\;\;\;\;\;\;\;\;\;\;v(0)=v_t,
\ee
where $(\rho_t,v_t)$ is the coordinate of the turning point of the extremal hypersurface in the bulk. With these boundary conditions,  
one has $H=\rho_t^{-n}$.
Given a particular form for $m(v)$, one may solve the  equations of motion \eqref{eom} to find the 
extremal hypersurface and thereby to compute the entanglement entropy. Of course, in general, it is not possible to solve the equations analytically, though
one may use a numerical method. 

It is worth noting that the model we are considering has three free parameters given by the 
dimension of the space-time, $D$, the scaling violating parameter $\theta$, and  the dynamical
exponent $z$. Therefore, one would naturally expect that the final results should depend on these three parameters. We note,
however, that for the entanglement entropy the results are just sensitive to the
effective dimension $d$ and the dynamical exponent $z$, up to an overall factor of $L^{D-1}$ which can be fixed by a dimensional analysis. In other words, even though we are dealing with 
a hyperscaling violating geometry,  nontrivial behaviors of the resultant entanglement entropy are 
the same as that of a Lifshitz geometry in $d+1$ dimensions. 

Actually, the holographic entanglement entropy 
for four and five dimensional Vaidya-Lifshitz solutions has been studied in Ref. \cite{Keranen:2011xs}. 
Thus, for effective dimensions $d=3$ and $d=4$ one can read
the results from that in \cite{Keranen:2011xs}. In particular, by making use of a numerical method it was shown that  the entanglement entropy grows  linearly with time and  then   saturates to its
equilibrium value at the saturation time  given by $t_s\sim \ell/2$.

In the next section, we shall further explore the behavior
of entanglement entropy (Wilson loop and  equal-time two-point function)
 for a global quantum quench holographically described by the time-dependent background
\eqref{solution2} with $Q(v)=0$ for the case of large entangling regions.


\section{GENERAL CONSIDERATION FOR LARGE ENTANGLING REGIONS}

 In this section, following the recent papers \cite{{Liu:2013iza},{Liu:2013qca}}, we will study 
entanglement entropy (Wilson loop and equal-time two-point function) for the case where
the size of the entangling region is large compared to the radius of horizon: $\ell/2 \gg \rho_H$. 
In this case, the system will reach  local equilibrium at $t\sim \rho_H$, which is earlier than  the time when
the entanglement entropy saturates to its equilibrium value. Indeed, as we will see, the evolution of 
 the extremal hypersurface for $t\gtrsim \rho_H$ is given by the geometry around and inside
the horizon. This is unlike the case of $\ell/2 \ll \rho_H$, where the entanglement entropy will be  saturated before  the time in which the system is locally equilibrated at $t\sim \rho_H$. Therefore, 
for small size entangling regions, there is no chance to probe the region inside the horizon. 
This is, indeed, our motivation to deal with large entangling regions.

Since we are interesting in a global quench in an uncharged system, in what follows, we will consider the following form for the function $f$, appearing in the $g_{vv}$ component of the metric \eqref{sol}:
\be
f(\rho,v)=1-\theta(v) (\frac{\rho}{\rho_H})^{d-1+z},
\ee
where $\theta(v)$ is the step function whose presence indicates that there is a rapid change in the
system as expected for a quench.

Now we would like to solve Eqs. \eqref{eom} for the function $f$ given  above. Since the function $f$ contains a step function, in order to solve the equations, it is
useful to study them in three separated regions: the $v<0$ region, the $v>0$ region, and the matching 
region at the null shell $v=0$.

{\bf (i) $v<0$ region.}

In this region, the step function is zero and thus $f(r,v)=1$. So the system is in 
the vacuum state whose gravity dual is given by  a hyperscaling violating solution as 
follows:
\be\label{I}
ds^2=\rho^{\frac{2(d-1)}{D}}\left(-\rho^{2-2z}dv^2-2\rho^{1-z}d\rho dv+d\vec{x}^2\right).
\ee
The scalar field and the gauge field are the same as that in Eq. \eqref{solution2}.

Since in this case $\frac{\partial f(\rho,v)}{\partial v}=0$ from the first equation in \eqref{eom}, one
finds that the momentum conjugate of $v$ is a constant of motion:
\be
P_{(i) v}=\rho^{1-z}(\rho'+\rho^{1-z} v')={\rm const}=0,
\ee
where the index $(i)$ denotes the value of quantities in the $v<0$ region. Here we have also imposed the boundary conditions \eqref{bdycondition} to show that the ``constant''
 is, indeed,  zero. On the other hand, using the conservation law \eqref{J}, one gets
\be\label{eq5}
v(\rho)=v_t+\frac{1}{z}(\rho_t^z-\rho^z),\hspace*{1cm}x(\rho)=\int_\rho^{\rho_t}\frac{d\xi \;\xi^{n}}{\sqrt{\rho_t^{2n}-\xi^{2n}}}.
\ee
Note  that at the null shell where $v=0$, from the above equation, one has
\be
\rho_c^z=\rho_t^z+z v_t
\ee
which gives the point where the extremal hypersurface intersects  the null shell, $\rho_c$. 
Moreover, by making use of Eq. \eqref{J},  one finds
\be
\rho_{(i)}'=-\rho_c^{1-z}v_{(i)}'=-\sqrt{\left(\frac{\rho_t}{\rho_c}\right)^{2n}-1}
\ee
where  the index $(i)$ denotes the value of quantities in the $v<0$ region.

{\bf (ii) $ v>0$ region.}

In this region the corresponding geometry is given by a static black hole  with a hyperscaling violating factor. More precisely, the corresponding geometry is
\be\label{II}
ds^2=\rho^{\frac{2(d-1)}{D}}\left(-\rho^{2-2z}\tilde{f}(\rho) dv^2-2\rho^{1-z}d\rho dv+d\vec{x}^2\right),\;\;\;{\rm with}\;\;
{\tilde f}(\rho)=1-(\frac{\rho}{\rho_H})^{d-1+z}\equiv 1-g(\rho),
\ee
and the other fields remain unchanged. 

In this case, again, $\frac{\partial f(\rho,v)}{\partial v}=0$, and therefore the momentum conjugate of $v$ is still a constant of motion, though its value is not zero:
\bea\label{Pvconst}
P_{(f)\;v}=\rho^{1-z}(\rho'+\rho^{1-z} v'{\tilde f}(\rho))={\rm const},
\eea
where  the index $(f)$ denotes the value of quantities in the $v>0$ region.

Plugging this equation into the conservation law \eqref{J}, one arrives at
\be\label{oneD}
\rho'^2=\frac{P_{(f) v}^2}{\rho^{2-2z}}+\left(\left(\frac{\rho_t}{\rho}\right)^{2n}-1\right){\tilde f}(\rho)\equiv V_{eff}(\rho),
\ee
which can be used to find
\be
\frac{dv}{d\rho}=-\frac{1}{\rho^{2(1-z)}{\tilde f}(\rho)}\left(\rho^{1-z}+\frac{P_{(f)v}}{\sqrt{V_{eff}(\rho)}}\right).
\ee
Here $V_{eff}(\rho)$  might be thought of as an effective potential for a one-dimensional 
dynamical system whose dynamical variable is $\rho$. In particular, the turning point of the
potential can be found by setting $V_{eff}(\rho)=0$. As we will see, such an interpretation has a physical 
impact in exploring  the behavior of entanglement entropy ( Wilson loop and equal-time two-point 
function).

{\bf (iii) Matching at the null shell.}

Having explored a possible solution of the equations \eqref{eom}  in $v<0$ and $v>0$ regions, it is crucial  to match the results of  these two regions at the null shell $v=0$. Of course, since $\rho$ and $v$ are 
the coordinates of the space-time they should be continuous across the null shell. We note, however, that,
since one is  injecting  matter along the null  direction $v$, one would expect that its corresponding 
momentum conjugate jumps once one moves from the $v<0$ region to that of $v>0$, whereas 
the momentum conjugate of $\rho$ must be continuous: $v'_{(f)}=v'_{(i)}$.
In fact, by integrating the equations of motion across the null shell and taking into account 
that the presence of a delta function would lead to a step function discontinuity after performing
the integration around $v=0$,  one  arrives at 
\be
\rho_{(f)}' =\left(1-\frac{1}{2}g(\rho_c)\right) \rho_{(i)}',\hspace*{1cm}{\cal L}_{(f)}={\cal L}_{(i)}.
\ee
It is, then, straightforward to read the momentum conjugate of $v$ in the $v>0$ region:
\bea\label{pfv}
P_{(f) v}=\frac{1}{2}\rho_c^{1-z}g(\rho_c)\rho_{(i)}'=-\frac{1}{2}\rho_c^{1-z}g(\rho_c)\sqrt{\left(\frac{\rho_t}{\rho_c}\right)^{2n}-1}.
\eea

Now we have all the ingredients to find the area of the corresponding extremal  hypersurface in the bulk.
In general, the extremal  hypersurface could be extended in both $v<0$ and  $v>0$ regions 
of the space-time. Therefore, the width $\ell$ and the boundary time could have contributions
from both regions:
\be\label{general1}
\frac{\ell}{2}=\rho_t\left(\int_{\frac{\rho_c}{\rho_t}}^{1}\frac{d\xi\;\xi^{n}}{\sqrt{1-\xi^{2n}}}+\int_0^{\frac{\rho_c}{\rho_t}}\frac{d\xi}{\sqrt{R(\xi)}}\right),\;\;\;\;\;
t=\rho_t^z\int_0^{\frac{\rho_c}{\rho_t}}\frac{d\xi\;\xi^{z-1}}{h(\xi)}\left(1+\frac{\xi^{z-1}E}
{\sqrt{R(\xi)}}\right),
\ee
where $E=P_{(f)v} \rho_t^{z-1}$, and 
\bea\label{def}
&&h(\xi)={\tilde f}(\rho_t\xi)=1-\left(\frac{\rho_t}{\rho_H}\right)^{d-1+z}\xi^{d-1+z},\cr &&\cr
&&R(\xi)=V_{eff}(\rho_t\xi)=E^2\xi^{2(z-1)}+\left(\frac{1}{\xi^{2n}}-1\right) h(\xi).
\eea
Finally, one finds
\be\label{general2}
{\cal A}_n=\frac{L^{\frac{D}{d-1}n-1}}{\rho_t^{n-1}}\left(\int_{\frac{\rho_c}{\rho_t}}^1\;\frac{d\xi}{\xi^{n}
\sqrt{1-\xi^{2n}}}
+
\int_{0}^{\frac{\rho_c}{\rho_t}}\frac{d\xi}{\xi^{2n}\sqrt{R(\xi)}}\right).
\ee
Note that, in general, ${\cal A}_n$ is divergent due to UV effects (large volume), and therefore it should be
 regularized by a UV cutoff.  Of course, since  in what follows we are mainly interested in the change of $\mathcal{A}_n$ when the system evolves from the vacuum state to an excited state,
 $\Delta \mathcal{A}_n= {\cal A}_n-{\cal A}_n^{\text{vac}}$, with
\bea\label{vac}
{\cal A}_n^{\text{vac}}=\frac{L^{\frac{D}{d-1}n-1}}{\rho_t^{n-1}}\int_{0}^1\;\frac{d\xi}{\xi^{n}
\sqrt{1-\xi^{2n}}},
\eea
the extremal hypersurface in the vacuum solution, ${\cal A}_n^{\text{vac}}$,  may be 
thought of as  a regulator.

In the rest of this section, using this general consideration we will study the behavior of 
entanglement entropy (Wilson loop and two-point function) during the process of thermalization after 
a global quench.



\subsection{Early time growth}

Let us study the behavior of entanglement entropy (Wilson loop and two-point function) 
 at early times when $t\ll \rho_H^z$. In this case, the crossing point, which is
the point  where the extremal hypersurface intersects the infalling shell, is very close to the boundary, so that $\frac{\rho_c}{\rho_H}\ll 1$. Therefore, one can expand the expressions of $t, \ell,$ and ${\cal A}_n$ in this limit. 
More precisely, from Eq. \eqref{general1} one finds
\bea\label{tearly}
t\approx \rho_t^z\int_0^{\frac{\rho_c}{\rho_t}}\; d\xi\;\frac{\xi^{z-1}}{h(\xi)}=\frac{\rho_c^z}{z}\left[1+\frac{z}{d-1+2z}\left(\frac{\rho_c}{\rho_H}\right)^{d-1+z}+\mathcal{O}\left(\left(\frac{\rho_c}{\rho_H}\right)^{2(d-1+z)}\right) 
\right],
\eea
while from Eq. \eqref{general2} and for $d+z-n\neq 0$ at leading order one gets
\be
{\cal A}_n\approx {\cal A}^{\rm vac}_n+ \frac{L^{\frac{D}{d-1}n-1}}{2(d+z-n)}\frac{\rho_c^{d+z-n}}{\rho_H^{d+z-1}}\left[1+{\cal O}\left(\left(\frac{\rho_c}{\rho_H}\right)^{d+z-1}\right)\right].
\ee
Here 
\be
 {\cal A}^{\rm vac}_n=L^{\frac{D}{d-1}n-1}\left(\frac{1}{(n-1)\epsilon^{n-1}}-\frac{c_n}{\ell^{n-1}}\right),
\ee
where $\epsilon$ is a UV cutoff and 
\be
c_n=\frac{2^{n-1}}{n-1}\left( \frac{\sqrt{\pi}\Gamma\left(\frac{1+n}{2n}\right)}{\Gamma\left(\frac{1}{2n}\right)}\right)^n.
\ee 
Therefore using Eq. \eqref{tearly} and setting $m=\rho_H^{1-d-z}$ at leading order one arrives at
\bea
\Delta {\cal A}_n\approx\frac{L^{\frac{D}{d-1}n-1}m}{2(d-n+z)}(zt)^{1+\frac{d-n}{z}},
\eea
which reduces to that of Vaidya-AdS for $z=1$\cite{{Liu:2013iza},{Liu:2013qca}}. For 
entanglement entropy where $n=d-1$, the above equation reads
\bea
\Delta {\cal A}_{d-1}\approx\frac{L^{D-1}m}{2(z+1)}(zt)^{1+\frac{1}{z}},
\eea
which is independent of $\theta$, as anticipated. So, even though we are dealing with a hyperscaling violating geometry,
the early time growth depends only on the dynamical exponent $z$ as if we had considered
a $D+2$-dimensional Lifshitz
geometry. It is also interesting to note that, for sufficiently large $z$ and arbitrary $n$, one finds a linear growth
at early times:
\bea\label{yu}
\Delta {\cal A}_n\approx\frac{L^{\frac{D}{d-1}n-1}m}{2 }\;t.
\eea

On the other hand, for $d+z-n=0$ from Eq. \eqref{general2} one gets a logarithmic behavior as follows\footnote{In fact, in Ref. \cite{serbyn:2013}, it was shown that in many-body strongly interacting disordered systems the entanglement entropy presents this universal slow growth behavior at early times. We thank Juan F. Pedraza  for a comment on this point.}:
\be
\Delta {\cal A}_n\approx \frac{L^{\frac{D}{d-1}n-1}m}{2}\;\ln \frac{\rho_c}{\rho_t}.
\ee
In this case, Eq. \eqref{tearly} is still valid, and therefore at leading order setting  $t=\rho_c^z/z$ one finds
\bea
\Delta {\cal A}_n\approx\frac{L^{\frac{D}{d-1}n-1}m}{2z}\ln \frac{z t}{\rho_t^z}.
\eea
It is important to note that, for the case of  entanglement entropy where $n=d-1$, the logarithmic behavior
occurs for $z=-1$ which  together with the null energy condition \eqref{NEC} requires leads to $\theta>D$.
We note, however, that  in this case the corresponding solution might be  unstable \cite{Dong:2012se}.
Of course, this is not the case  we are interested in (see Eq. \eqref{NEC} and a line after). Nevertheless, for Wilson loop and 
equal-time two-point function there is a possibility to have such a behavior, while we are within the
range of interest.


\subsection{Growth in the intermediate region}

In this subsection, we will consider the intermediate region when $\rho_H^z\lesssim t\lesssim \rho_H^{z-1}\frac{\ell}{2}$. 
In this case a crucial observation which has been made in Refs. \cite{{Liu:2013iza},{Liu:2013qca}}
 is as follows
(see also \cite{Hubeny:2013dea} for further discussions).

Actually, in this 
case the hypersurface could penetrate inside the horizon, typically intersects the null shell
at $\rho_c> \rho_H$, and reaches the turning point at $\rho_t>\rho_c$. Moreover, there is a ``critical'' extremal 
hypersurface which intersects the null shell behind the horizon at a critical point $\rho^*_c$.
Those hypersurfaces which intersect the null shell at $\rho_c< \rho^*_c$ will reach the
boundary, while those that intersect at  $\rho_c> \rho^*_c$ never reach the boundary and indeed fall
into the singularity. 

To study the critical  extremal hypersurface, we note that  Eq. \eqref{oneD} may be 
considered as the energy conservation law for a one-dimensional dynamical system whose
effective potential is given by $V_{eff}(\rho)$. Stable trajectories may occur around the 
minimum of the potential. Indeed, one may consider a special case where at the minimum
both the velocity and the acceleration are zero. In this case, the particle remains fixed at this point.
Of course, generally,  it is not obvious whether such a point exists.

Actually, for a fixed  turning point $\rho_t$, there is a free parameter in the effective potential
given by $\rho_c$ which 
may be tuned to a particular value $\rho_c=\rho^*_c$ such that the minimum of the 
effective potential becomes zero. In other words, one may have
\bea\label{criticalhyper}
\frac{\partial V_{eff}(\rho)}{\partial\rho}\bigg|_{\rho_m,\rho^*_c}=0,\;\;\;\;\;\;\;\;\;\;\;\;
V_{eff}(\rho)|_{\rho_m,\rho^*_c}=0.
\eea
Therefore, if the hypersurface intersects the null shell at the critical point, it remains fixed at 
$\rho_m$. Here $\rho_m$ is a point which minimizes the effective potential.
This is, indeed, the critical extremal hypersurface which is responsible for the 
linear growth in the intermediate region as we will see below. 

To compute the width $\ell$, time $t$, and ${\cal A}_n$ given in Eqs. \eqref{general1} and \eqref{general2} around the
critical extremal hypersurface, we will consider $\rho_c=\rho^*_c (1-\delta)$ for $\delta\ll 1$.
In this limit the main contributions to the integrals in \eqref{general1} and \eqref{general2} 
come from the $\rho\sim \rho_m$ region, where we are close to the minimum of the  effective potential.
In this limit,  the dominant term in Eqs. \eqref{general1} and \eqref{general2} is the term which contains the $\frac{1}{\sqrt{R(\xi)}}$ factor that develops a single pole
singularity.  More precisely, for $\rho_c=\rho^*_c (1-\delta)$ and near
$\xi\sim \frac{\rho_m}{\rho_t} \equiv \xi_m$ one  gets
\bea
R(\xi)=b\delta+\frac{1}{2}(\xi-\xi_m)^2R''(\xi_m)+... ,
\eea
where  $b=-\xi_c^*\frac{dR}{d\xi_c}\big|_{\xi_c=\xi_c^*}$. Therefore one arrives at 
\bea
t &\approx&
\rho_t^z\int_{\xi\sim\xi_m}\; d\xi\frac{\xi_m^{2(z-1)}E^*}{h(\xi_m)\sqrt{b\delta+\frac{1}{2}R''(\xi_m)(\xi-\xi_m)^2}}=
-\rho_t^z\frac{\xi_m^{2(z-1)}E^*}{h(\xi_m)\sqrt{\frac{1}{2}R''(\xi_m)}}\ln\delta,\nonumber\\
\frac{\ell}{2}&\approx &
b_n\rho_t+\rho_t\int_{\xi\sim\xi_m}\;\frac{d\xi}{h(\xi_m)\sqrt{b\delta+\frac{1}{2}R''(\xi_m)(\xi-\xi_m)^2}}=
b_n\rho_t-\frac{\rho_t}{\sqrt{\frac{1}{2}R''(\xi_m)}}\ln\delta,
\eea
with $b_{n}=\frac{\sqrt{\pi}\Gamma(\frac{1}{2}+\frac{1}{2n})}{\Gamma(\frac{1}{2n})}$ and
\be
R''(\xi_m)=\frac{\partial^2 R(\xi)}{\partial \xi^2}\bigg|_{\xi_m,\rho^*_c}
=\rho_t^2\frac{\partial^2 R(\rho)}{\partial \rho^2}\bigg|_{\rho_m,\rho^*_c}, \;\;\;\;\;\;\;\;\;E^*=
-\left(\frac{\rho_t}{\rho_m}\right)^{z-1}\sqrt{-h(\xi_m)\left(\left(\frac{\rho_t}
{\rho_m}\right)^{2n}-1\right)}.
\ee
Here the functions $R(\xi)$ and $h(\xi)$ are those defined in Eq. \eqref{def}. On the other hand, from Eqs. \eqref{general2} and \eqref{vac}, one finds
\be
{\cal A}_n \approx  {\cal A}^{\rm vac}_n+\frac{L^{\frac{D}{d-1}n-1}}{\rho_t^{n-1}} \int_{\xi\sim\xi_m}\; \frac{d\xi}{\xi_m^{2n}\sqrt{b\delta+\frac{1}{2}R''(\xi_m)
(\xi-\xi_m)^2}}={\cal A}^{\rm vac}_n
-\frac{L^{\frac{D}{d-1}n-1}}{\rho_t^{n-1}}  \frac{1}{\xi_m^{2n}\sqrt{\frac{1}{2}R''(\xi_m)}}\ln\delta,
\ee
that leads to the following linear growth:
\be
\Delta {\cal A}_n \approx \frac{L^{\frac{D}{d-1}n-1}}{\rho_t^{n+z-1}}\;\frac{h(\xi_m)}{\xi_m^{2( n+z-1)}E^*}
\; t,
\ee
which in the large $\rho_t$ limit  reads
\be\label{linear}
\Delta {\cal A}_n \approx L^{\frac{D}{d-1}n-1}\;\frac{v_n t}{\rho_H^{ n+z-1}},
\ee
where 
\be
v_n=\sqrt{-{\tilde f}(\rho_m)}\;(\frac{\rho_H}{\rho_m})^{n+z-1}.
\ee 
As we will see,  $v_n$ is a numerical factor which is independent of the shape of the entangling region but depends on the final equilibrium state. This parameter might be thought of as the velocity of the linear growth.


\subsection{Late time saturation}

In general, the  late time behavior of the entanglement entropy (Wilson loop and equal-time two-point function) after a  global quench 
depends on  details of the system as well as the shape of the entangling region.  Nevertheless, one would expect that, if one waits enough, the entanglement entropy  saturates to
its equilibrium value which is essentially that of  a thermal state. In this subsection, using Eq. \eqref{general2} we show
how this happens.

To proceed, we note that for time $t\gtrsim \rho_H$ the system is locally equilibrated and there is
a saturation time after which the extremal hypersurface is entirely outside the horizon: $\rho_t<\rho_H$.
On the other hand, since we are interested in large entangling regions, the main contribution 
to ${\cal A}_n$ comes from the geometry around the horizon. Actually, in the present case 
in order  to compute the width $\ell$ and ${\cal A}_n$ one should expand Eqs. \eqref{general1} and
\eqref{general2} for $\rho_c\approx \rho_t\rightarrow \rho_H$. Note that in this limit  $P_{(f)v}\approx 0$. Therefore setting $\rho_c=\rho_t (1-\delta)$ for $\delta\ll 1$ and $\rho_t\approx \rho_H$
one finds
\bea\label{ww}
\frac{\ell}{2} \approx  \rho_H\int_0^{1-\delta}\frac{d\xi}{\sqrt{R(\xi)}},\;\;\;\;\;\;\;\;\;\;\;
 {\cal A}_n \approx  \frac{L^{\frac{D}{d-1}n-1}}{\rho_H^{n-1}}
\int_{\frac{\epsilon}{\rho_H}}^{1-\delta}\frac{d\xi}{\xi^{2n}\sqrt{R(\xi)}},
\eea
where $\epsilon$ is a UV cutoff. It is worth noting that in this limit, apart from the UV divergence of ${\cal A}_n$, which is due to a double
zero at $\xi=1$  in the square root,  the main contributions to the width $\ell$ and ${\cal A}_n$
come from the $\xi=1$ point. Around this point, one may recast ${\cal A}_n$ to the
following form:
\be
  {\cal A}_n \approx  \frac{L^{\frac{D}{d-1}n-1}}{\rho_H^{n-1}}\left(
\int_0^{1-\delta}\frac{d\xi}{\sqrt{R(\xi)}}+\int_{\frac{\epsilon}{\rho_H}}^{1}d\xi\;\frac{(1-\xi^{2n})}{\xi^{2n}\sqrt{R(\xi)}}
\right).
\ee
One observes that the first term is exactly the one which appears in the expression of $\ell$, and  moreover
the second term is finite at $\xi=1$ while it diverges at the UV limit which  is regularized by the
cutoff $\epsilon$. From these observations, one finds 
\be\label{sat}
{\cal A}^{\rm sat}_n\approx \frac{L^{\frac{D}{d-1}n-1}}{(n-1)\epsilon^{n-1}}+\frac{L^{\frac{D}{d-1}n-1}\ell}{2\rho_H^{n}}+\cdots\ ,
\ee
where ${\cal A}_n^{\rm sat}$ denotes the equilibrium value of ${\cal A}_n$. Note that for the entanglement entropy where $n=d-1$ it is indeed  the same  as that in the static black hole 
given in Eq. \eqref{En-thermal}, as expected.

It is also possible to estimate the saturation time. To do so, it is, however, important to note that,
as we have already mentioned, the  nature of saturation (and therefore the saturation time)
depends on the shape of the entangling 
region as well as  the parameters of the model such as the effective
dimension $d$ and  the dynamical exponent $z$. In particular, saturation could occur continuously or  discontinuously. In the discontinuous case, although ${\cal A}_n$ is
continuous at $t_s$, its first derivative is  discontinuous\cite{Liu:2013qca}. 
In the continuous case, the saturation is continuous, and then the saturation time may be 
calculated from Eq. \eqref{general1} in the limit of $\rho_c=\rho_t(1-\delta)$ for 
$\rho_t\approx \rho_H$. In this limit we find
\bea\label{tsatur}
\frac{\ell}{2} \approx  \rho_H\int_0^{1-\delta}\frac{d\xi}{\sqrt{R(\xi)}}\approx -\frac{\rho_H}{\sqrt{2n(d+z-1)}}\ln\delta,\;\;\;\;\;
t_s\approx \rho_H^z \int_0^{1-\delta}\frac{d\xi \;\xi^{z-1}}{h(\xi)}\approx -\frac{\rho_H}{d+z-1}\ln\delta
\eea
which results to the following saturation time:
\be
t_s\approx\rho_H^{z-1}\sqrt{\frac{2n}{d+z-1}}\;\frac{\ell}{2}.
\ee
Although in the case of discontinuous saturation there is no general formula for $t_s$, one may still  find a distinctive  time scale which  could be thought of as 
a characteristic time for saturation.  
Indeed, assuming to have the linear growth all the way to the saturation point, one may find the 
distinctive  time by equating \eqref{linear} and \eqref{sat}. Doing so, one finds
\be
-\frac{c_0}{\ell^{n-1}}+\frac{v_nt_l}{\rho_H^{n+z-1}}\approx \frac{\ell}{2\rho_H^{n}},
\ee
which can be solved to find a characteristic time for saturation  as follows:
\be
v_nt_l\approx \rho_H^{z-1}\;\frac{\ell}{2}+c_0\frac{\rho_H^{n+z-1}}{\ell^{n-1}}.
\ee
Since we are interested in $\ell\gg \rho_H$, the saturation time is in fact $t_l\sim \frac{\rho_H^{z-1}}{v_n}\frac{\ell}{2}$. Note that, in general, $t_l>t_s$. It is then evident that in the case of continuous
saturation the linear growth cannot persist all the way to the saturation point, and the
saturation time is given by $t_s$.

To see whether 
the saturation is continuous, one may look at $t-t_s$ in the limit of $\rho_c\rightarrow\rho_t$.
In fact, saturation is continuous if $t-t_s$ becomes negative in the $\rho_t-\rho_c\rightarrow
0$ limit\cite{Liu:2013qca}.  
Actually, by making use of the definition of $t_s$ and Eq. \eqref{general1}, it is straightforward to compute $t-t_s$ (the details of the following computations are summarized in Appendix B):
\bea\label{tts}
t-t_s=a\sqrt{2n \;\delta}+{\cal O}(\delta^2),\;\;\;\;\;\;\;\;\;\;\;{\rm with}\;\;a=\frac{\rho_t^z}{2} g(\rho_t)\left(\frac{1}{n \tilde{f}^2(\rho_t) F'(\rho_t)}-
H(\rho_t)\right).
\eea
Here
\bea\label{FH}
F(\rho_t)=\rho_t\int_0^1\frac{\xi^n d\xi}{\sqrt{h(\xi)(1-\xi^{2n})}},\;\;\;\;\;\;\;
H(\rho_t)=\int_0^1\frac{\xi^{2(z-1)+n} d\xi}{\sqrt{h^3(\xi)(1-\xi^{2n})}}.
\eea
where $h(\xi)$ is given in Eq. \eqref{def}. Saturation is continuous if $a<0$ and
discontinuous if $a>0$. One can then look at the sign of $a$ to see whether the saturation
is continuous. Actually, in the present case where we are considering the entanglement entropy
for a strip, the saturation is discontinuous, and therefore $t_l$ gives a rough estimation 
of the saturation time. Indeed, the situation is the same as that of the AdS-Vaidya metric\cite{Liu:2013qca}.

\subsection{More details for entanglement entropy}

In this subsection, setting $n=d-1$ we will present  explicit values of the parameters we 
have considered in the previous subsections. Starting from the effective potential, one 
finds that the effective potential is minimized 
at $\rho_m$, which can be obtained from the following equation:
\be\label{eq1}
\rho_t^{2(d-1)} =\rho_m^{2(d-1)}\frac{ 2
\rho_m \tilde{f}'(\rho_m)+(z-1)g^2(\rho_c)(\frac{\rho_m}{\rho_c})^{2( z-1)}}{
2\rho_m \tilde{f}'(\rho_m)-4 (d-1) \tilde{f}(\rho_m)+
(z-1) g^2(\rho_c) \left(\frac{\rho_m}{\rho_c}\right)^{2 (d-2+z)} }.
\ee
Note that, unlike the $z=1$ case, for a fixed horizon radius the value of the radial coordinate at  which 
the effective potential is minimized, $\rho_m$, is both a function of the extremal hypersurface
turning point in the bulk and the point 
where the hypersurface intersects the null shell. Therefore the turning point would not completely fix $\rho_m$. 

We note, however, that for the critical extremal hypersurface defined in \eqref{criticalhyper} it is possible to fix both $\rho_m$ and $\rho^*_c$. Indeed, for 
the critical point we require that the effective potential is also zero at the minimum point. So, one finds
\be\label{eq2}
\rho_t^{2(d-1)}=\rho_m^{2(d-1)}\frac{4 \tilde{f}(\rho_m)+g^2(\rho^*_c) (\frac{\rho_m}{\rho^*_c})^{2 (z-1)}}
{4 \tilde{f}(\rho_m)+g^2(\rho^*_c) (\frac{\rho_m}{\rho^*_c})^{2 (d-2+z)}}. 
\ee
This may also be considered as another relation which fixes  $\rho_m$ as a function of $\rho_t$ at
the critical point. Therefore, solving Eqs. \eqref{eq1} and \eqref{eq2} together one can 
find $\rho^*_c$ and the corresponding $\rho_m$ of the critical extremal hypersurface. Indeed, as 
far as  our considerations in the previous subsection are concerned, these are what we need to proceed 
 exploring the behavior of the entanglement entropy in the intermediate region.
In particular, in the large $\rho_t$ limit, assuming $\rho_m$ and 
$\rho_c^*$ remain finite,  one gets
\be\label{ii}
\frac{\rho_m}{\rho_H}=\left(\frac{2(d+z-2)}{d+z-3}\right)^{\frac{1}{d+z-1}},\;\;\;\;\;\;\;\;\;
\frac{\rho^*_c}{\rho_H}=2\sqrt{\frac{d+z-1}{d+z-3}}\left(\frac{d+z-3}{2(d+z-2)}\right)^{\frac{d+z-2}{d+z-1}}.
\ee
It is worth noting  that from the null energy condition \eqref{NEC} one has $d>1, z>1$, leading to
$d+z>2$. Therefore, the above expressions do not make sense for $d+z<3$.\footnote{
Actually, for $d+z<3$ it is not possible to keep both $\rho_m$ and $\rho_c^*$ finite
 for large $\rho_t$. Indeed, in this case the entanglement entropy  does not have 
a linear growth in the intermediate region.}
Thus, one could consider only the case where $d+z\geq3$. In particular, for $d+z>3$, where the above
expressions are well defined, 
the entanglement entropy of the 
system exhibits a linear growth in the intermediate region with the following velocity:
\be
v_E=v_{d-1}=\left(\frac{d+z-3}{2(d+z-2)}\right)^{\frac{d+z-2}{d+z-1}}\sqrt{\frac{d+z-1}{d+z-3}}.
\ee 

On the other hand, for  $d+z=3$, although for large $\rho_t$, the critical point $\rho_c^*$ remains
finite, $\rho_m$ becomes large and therefore the expressions in Eq. \eqref{ii} are not valid.
In fact, in the present case one  arrives at
\be
\rho_m=\sqrt{\rho_H\rho_t},\;\;\;\;\;\;\;\;\;\rho_c^*=2\rho_H.
\ee
In this case, the entanglement 
entropy still has a linear growth in the intermediate region with $v_E=1$.  Note that, 
even though one could be in an arbitrary dimension, the situation is very similar 
to that of two-dimensional quantum quench, where the velocity is $v_E=1$ and  the entanglement
entropy  saturates at $
t_s\sim \frac{\ell}{2}$\cite{CC}.


\section{CONCLUSIONS}

In this paper, we have considered an Einstein-Maxwell-dilaton theory with a nontrivial potential for the dilaton. We have  obtained an analytic solution with a form of Vaidya-charged black hole solution 
with a hyperscaling violating factor.  This solution may be thought of as a model describing 
gravitational collapse of charged matter to
make a charged black hole with a hyperscaling violating factor. 

From the gauge/gravity duality point of view, this geometry may provide a holographic description for a 
global quantum quench for a strongly coupled field theory with hyperscaling violation and an anisotropic 
scaling symmetry.
The quantum quench which might be caused by a  rapid change in the theory would correspond 
to an instant injection of matter in a small time interval $\delta t$.

This system may also be used to examine the process of the thermalization in the model after a global quantum quench. Therefore,  in order to probe the thermalization caused by the global quantum quench we have studied the time dependence of entanglement entropy ( Wilson loop and equal-time two-point function). Holographically, this can be done  by  extremizing a certain codimension-two  hypersurface  in the bulk
geometry \eqref{solution2}.  Although we have mainly considered the entanglement entropy, we have 
worked in a setup so that the final results could be extended to the Wilson loop and 
equal-time two-point function of an operator with a large conformal dimension by setting $n=2\frac{d-1}{D}$, and $n=\frac{d-1}{D}$, respectively. 


In this paper, following Refs. \cite{{Liu:2013iza},{Liu:2013qca}} we have considered 
the case where $\frac{\ell}{2}\gg \rho_H$ and therefore the evaluation of the
corresponding hypersurface  is controlled by regions inside and around the horizon. 
We have found that at early times the growth of the entanglement entropy depends on the dynamical exponent $z$, which indicates that entanglement entropy at early times is sensitive to the
initial  state, while in the intermediate region it always grows linearly.  We have, however, observed that, 
in the large $z$ limit, the early time behavior is universal ( in the sense that 
it is independent of $n$)  and it grows linearly.

For the interesting case of $d=2$ where the dual theory exhibits a Fermi surface\cite{Ogawa:2011bz},
and for an arbitrary $z> 1$, the
velocity reads
\be
\frac{1}{2}\leq v_E=\left(\frac{z-1}{2z}\right)^{\frac{z}{z+1}}\sqrt{\frac{z+1}{z-1}}< 1.
\ee 
Here one may reach $v_E=\frac{1}{2}$ in the large $z$ limit where we have linear growth all the way from the initial
phase up to the saturation phase.
 The velocity is sensitive to the dynamical exponent  which in turns shows that 
the velocity (the growth) depends on the initial state.
Note also that for $d\geq 3$ and arbitrary $z$ the 
velocity is always less than one. 
It would be interesting to see whether this  behavior of velocity may be understood by
a free-streaming model\cite{CC}.

Following Ref. \cite{Liu:2013iza}, one may define a dimensionless rate of growth as follows:
\be
{\cal R}(t)=\frac{1}{L^{\frac{D}{d-1}n-1}\mathcal{A}_n^{(f)} \rho_H^{1-z}}\;\frac{\partial S}{\partial t},
\ee
which, in the intermediate region where one has a linear growth, is equal to the velocity of evolution 
growth that  is always less than one. 

Although in this paper we have considered only the strip case, 
its generalization to a sphere should be straightforward.  Moreover, it would also be interesting 
to study the model for the case where the background is charged: $Q(v)\neq 0$.

  
\section*{ACKNOWLEDGEMENTS}

We thank  Mohammad Reza Tanhayi for related discussions. M.R.M.M. also thanks Ali Mollabashi for useful discussions. We also thank J.~F.~Pedraza for his useful comments.
This work is supported by Iran National Science Foundation (INSF).

\section*{APPENDIX}
\appendix
\section{ENTANGLEMENT ENTROPY FOR STATIC SOLUTION} \label{App:AppendixA}

In  this Appendix  we will review certain properties of the entanglement entropy of a strongly coupled field theory whose gravitational description is given by the 
background \eqref{solution} (for details, see, for example, 
\cite{{Dong:2012se},{Alishahiha:2012cm},{alishahiha:2012}}). To compute the entanglement entropy via AdS/CFT correspondence,
one needs to minimize a surface in the bulk gravity.  More precisely, given a gravitational  theory with
 the  bulk Newton constant $G_N$, the holographic entanglement entropy 
is given by \cite{RT:2006PRL,RT:2006}
\be\label{EE}
S_A=\frac{\mathrm{area}(\gamma_A)}{4G_N},
\ee
where $\gamma_A$ is the  minimal surface in the bulk whose boundary coincides with the boundary of the entangling region. 

To proceed let us consider a long strip in the dual theory given by
\be
t={\rm fixed},\;\;\;\;\;\;\;\;-\frac{\ell}{2}\leq x_1\leq \frac{\ell}{2},\;\;\;\;\;\;\;0\leq x_i\leq L\;\;\;\;\;\;{\rm for}\;i=2,\cdots,D.
\ee
The codimension-two hypersurface  $\gamma_A$ in the bulk may be parametrized by $x_1=x(r)$, so that
the induced metric on this hypersurface, setting $r=\frac{1}{\rho}$, is 
\be
ds^2=\rho^{2\frac{\theta}{D}-2}\bigg[\left(\frac{1}{f(\rho)}+x'^2\right)d\rho^2+d\vec{x}^2\bigg].
\ee
Therefore, the area $A$ reads
\be\label{A0}
{\cal A}=\frac{L^{D-1}}{2}\int d\rho \frac{\sqrt{{f}^{-1}+x'^2}}{\rho^{d-1}},\;\;\;\;\;\;{\rm with}\;\;\;
f(\rho)=1-{m}{\rho^{d-1+z}}+{Q^2}{\rho^{2(d-2+z)}}.
\ee
where a prime represents derivative with respect to $\rho$. It is then straightforward to minimize the above area to arrive at
\be
\frac{\ell}{2}=\int_0^{\rho_t}d\rho \frac{\left(\frac{\rho}{\rho_t}\right)^{d-1}}{
\sqrt{f(\rho)\left(1-\left(\frac{\rho}{\rho_t}\right)^{2(d-1)}\right)}},\;\;\;\;\;\;\;\;\;
S=\frac{L^{D-1}}{4G_N}\int_\epsilon^{\rho_t}d\rho \frac{1}{\rho^{d-1}
\sqrt{f(\rho) \left(1-\left(\frac{\rho}{\rho_t}\right)^{2(d-1)}\right)}}
\ee
where $\rho_t$ is the extremal hypersurface
turning point in the bulk and $\epsilon$ is a UV cutoff.

 For 
$f=1$, which corresponds to a vacuum solution, one finds\cite{Dong:2012se}
\be
S_{\text{vac}}=\left\{\ba{ll} 
\frac{L^{D-1}}{4G_N}\left(\frac{1}{(d-2)\epsilon^{d-2}}-\frac{c_{d-1}}{\ell^{d-2}}\right)
&{\rm for}\;\;d\neq 2,\cr &\cr
\frac{L^{D-1}}{4G_N}\ln \frac{\ell}{\epsilon},&  {\rm for}\;\;d= 2.
\ea \right.
\ee

On the other hand, for an excited state whose gravitational dual is provided by the black brane solution
\eqref{solution}, the corresponding entanglement entropy may be found by minimizing the area 
when $f\neq 1$. In this case, in general, it is not possible to find an explicit expression for 
the  entanglement entropy, though in certain limits one may extract the general behavior of 
the entanglement entropy.

Actually, for sufficiently small entangling regions, it is possible 
to expand the area expression \eqref{A0} to find the change of area due to the change of the 
geometry. More precisely, for $m\ell\ll 1$ one has
\be
\Delta {\cal A}=\frac{L^{D-1}}{2}\int d\rho\;\; \delta_f\left(\frac{\sqrt{{f}^{-1}+x'^2}}{\rho^{d-1}}\right)\bigg|_{f=1}\Delta f,
\ee
 which leads to the following expression for the change of the entropy:
\be
\Delta S=S-S_{\text{vac}}=\frac{L^{D-1}\ell^{z}}{16G_N(d-2)}\left(c_m m \ell-c_Q Q^2 \ell^{d-2+z}\right),
\ee
where $c_m=c^z$ and $c_Q=c^{d+2z-3}$ with $c=\Gamma(\frac{1}{2(d-1)})/(2\sqrt{\pi}\Gamma(\frac{d}{2(d-1)}))$. Note that, upon the identification of entanglement temperature with $\ell$, as $T_{E}\sim \ell^{-z}$ the above expression may be identified as the first 
law of entanglement entropy\cite{{Bhattacharya:2012mi},{Allahbakhshi:2013rda},{Blanco:2013joa},{Wong:2013gua}}; see also \cite{Pal:2013fha} for the $z=1$ case.

For sufficiently large entangling regions, the main contributions come from the limit where the minimal surface
is extended all the way to the  horizon so that $\rho_t\sim \rho_H$ and then 
\be\label{En-thermal}
 \Delta S=\frac{L^{D-1} \ell}{8G_N\rho_{H}^{d-1}}=L^{D-1}\frac{\ell}{2} {\cal S}_{\text{BH}},
\ee
where ${\cal S}_{BH}$ is the density of the thermal entropy given in Eq. \eqref{thermal}.

\section{DETAILS OF COMPUTATIONS FOR SATURATION TIME} \label{App:AppendixB}

Here we briefly review the behavior of the extremal hypersurface near the saturation point. As mentioned before, the computations are similar to Ref. \cite{{Liu:2013qca}}; we just redo them with an emphasis on the role of $z$ and $\theta$. Near the saturation point, both $\rho_c$ and $\rho_t$ tends to $\rho_b$ (i.e., the turning point of the extremal hypersurface in the static black brane background), so
\bea\label{rbrcrt}
\rho_c=\rho_t\left(1-\delta\right),\;\;\;\;\;\;\;\;\rho_t=\rho_b\left(1+\epsilon\right),
\eea
where $\epsilon, \delta \ll 1$. Now using \eqref{pfv} one arrives
\bea\label{Esat}
E=-\frac{1}{2}g(\rho_t)\sqrt{2n\delta}+...\;.
\eea
In order to find the relation between these parameters setting $\rho_c=\rho_t=\rho_b$ and $P_{(f)v}=0$ in \eqref{general1}, one finds
\bea
\frac{\ell}{2}=\rho_b\int_0^1\frac{d\xi\;\xi^n}{\sqrt{\left(1-\xi^{2n}\right) h(\xi)}}\equiv F(\rho_b).
\eea
Using this definition leads to
\bea
F(\rho_b)-F(\rho_t)=\rho_t\int_{\frac{\rho_c}{\rho_t}}^{1}\frac{d\xi \;\xi^{n}}{\sqrt{1-\xi^{2n}}}-\rho_t\int_{\frac{\rho_c}{\rho_t}}^{1}\frac{d\xi}{\sqrt{R(\xi)}}+\rho_t\int_{0}^{1}d\xi\left(\frac{1}{\sqrt{R(\xi)}}-\frac{\xi^n}{\sqrt{(1-\xi^{2n})h(\xi)}}\right).
\eea
At leading order the first two terms have the same contribution $\frac{\rho_t}{n}\sqrt{2n\delta}$, but with different signs, and therefore cancel each other. The contribution of the third term is $\frac{-\rho_tg(\rho_t)}{2nh(1)}\sqrt{2n\delta}$, leading to
\bea
F(\rho_b)-F(\rho_t)=\frac{g(\rho_t)}{h(1)}\frac{1}{\epsilon}\sqrt{\frac{\delta}{2n}},
\eea
which after using \eqref{rbrcrt} becomes
\bea\label{epsidelta}
\epsilon=\frac{g(\rho_t)}{h(1)F'(\rho_t)}\sqrt{\frac{\delta}{2n}}.
\eea
Now by considering the definition of $t_s$
\bea
t_s=\rho_b^z\int_0^1\frac{d\xi\;\xi^{z-1}}{h(\xi)\big|_{\rho_t=\rho_b}},
\eea
one can rewrite \eqref{general1} as
\bea\label{ttsappendix}
t-t_s=\rho_t^z\int_{\frac{\rho_b}{\rho_t}}^{\frac{\rho_c}{\rho_t}}\frac{d\xi\;\xi^{z-1}}{h(\xi)}+\rho_t^z\int_0^{1}\frac{d\xi \;\xi^{2(z-1)}E}{h(\xi)\sqrt{R(\xi)}}-\rho_t^z\int_{\frac{\rho_c}{\rho_t}}^{1}\frac{d\xi \;\xi^{2(z-1)}E}{h(\xi)\sqrt{R(\xi)}}.
\eea
The contributions of the first two terms are $ \frac{\rho_t^z}{h(1)}\epsilon$ and $\rho_t^z E H(\rho_t)$, respectively, where $H(\rho_t)$ is defined in \eqref{FH}. The third term is of the order of $\mathcal{O}(\epsilon^2)$ and we neglect it. Finally, plugging these expressions into \eqref{ttsappendix} and using \eqref{Esat} and \eqref{epsidelta} one can find \eqref{tts}.

\end{document}